\documentclass[preprint,12pt]{elsarticle}

\usepackage{amssymb}
\usepackage{xcolor}

\journal{NIMA}

\newcommand{\B}{ {\bf {B}}}

\newcommand{\p}{ {\bf {p}}}

\newcommand{\x}{ {\bf {x}}}
\newcommand{\vv}{ {\bf {v}}}
\newcommand{\nhat}{ {\bf {\hat{n}}}}
\begin{document}

\begin{frontmatter}



\title{ Stochastic modeling and survival analysis of marginally trapped neutrons
for a magnetic trapping neutron lifetime experiment}






\author[NISTB]{K.\,J. Coakley\corref{kjc}}
\ead{kevin\.coakley@nist.gov}
\author[NISTG]{M.\,S. Dewey}
\author[NISTG]{M.\,G. Huber}
\author[NCSU,TUNL]{P.\,R. Huffman}
\author[NCSU,TUNL]{C.\,R. Huffer}
\author[NISTG,NCSU]{D. E. Marley}
\author[NISTG]{H.\,P. Mumm}
\author[UNC,TUNL]{C.\,M. O'Shaughnessy}
\author[NCSU,TUNL]{K.\,W. Schelhammer}
\author[NISTG]{A.\,K. Thompson}
\author[NISTG]{A.\,T. Yue}

\address[NISTB]{National Institute of Standards and Technology, 325
Broadway, Boulder, CO 80305}
\address[NISTG]{National Institute of Standards and Technology, 100
Bureau Drive, Stop 8461, Gaithersburg, MD 20899}
\address[NCSU]{North Carolina State University, 2401 Stinson Drive,
Box 8202, Raleigh, NC 27695}
\address[TUNL]{Triangle Universities Nuclear Laboratory, 116 Science
Drive, Box 90308, Durham, NC 27708}
\address[UNC]{University of North Carolina at Chapel Hill, 120 E.
Cameron Ave., CB \#3255, Chapel Hill, NC 27599}
\cortext[kjc]{Corresponding author}

\begin{abstract}
In a variety of neutron lifetime experiments,
in addition to $\beta-$decay,
neutrons can be lost by 
other mechanisms including wall losses.
Failure to account for these
other loss mechanisms produces
systematic measurement error
and associated systematic uncertainties in
neutron lifetime measurements.
In this work, we construct a competing risks
survival analysis model to account for
losses due to the joint effect of $\beta-$decay losses,
wall losses of
marginally  trapped neutrons, and an additional absorption mechanism.
We
determine the
survival probability function associated
with the wall loss mechanism 
by a Monte Carlo method.
We
track
marginally trapped
neutrons with a symplectic
integration method that assumes neutrons
are classical particles.
We model wall loss probabilities
of Ultracold Neutrons (UCNs) that collide with trap boundaries
with a quantum optical model.
Based on a
fit of the competing risks model
to a subset of
the NIST experimental data,
we determine the mean lifetime of trapped neutrons
to be
approximately 700 s
-- considerably less than the
current best estimate 
of (880.1 $\pm$ 1.1) s
promulgated by the Particle Data Group [1].
Currently,
experimental studies are underway to determine if
this discrepancy can be explained by
neutron capture by
${}^3$He impurities in the trapping volume.
We also quantify uncertainties
associated with Monte Carlo sampling
variability and imperfect knowledge of
physical models for neutron interactions with
materials at the walls of trap
as well
as beam divergence
effects. 
Finally,
in a Monte Carlo experiment,
we demonstrate that
when the trapping potential is ramped down and
then back up again,
systematic error due to wall losses of marginally trapped 
neutrons
can be suppressed to a
very low level.
Monte Carlo
simulation studies indicate that
this ramping strategy is more efficient than 
an alternative ramping scheme
where UCNs are produced when the field is fully ramped,
and then increased to its maximum value after the trap
is filled. 
The survival probability formalism developed here
should be applicable
to other experiments where
neutron (or other particle)
loss mechanisms
are non-trivial (i.e., where the associated
survival probability function with the loss mechanism
is non-exponential).
\end{abstract}


\end{frontmatter}



\section{Introduction}
According to the
Particle Data Group,
the current best estimate of the mean lifetime of the neutron 
$\tau_n$
is
(880.1 $\pm$ 1.1) s [1].
This uncertainty is largely due to
systematic effects. Hence, reduction of systematic
uncertainties is key to reducing the
overall uncertainty
of the mean lifetime of the neutron [2].
To emphasize this point, we note that
the difference between
neutron lifetime estimates determined by
beam and bottle experiments 
is (8.4 $\pm$ 2.2) s. This difference
is likely due to unresolved systematic errors
[3].
In trapping experiments,
marginally trapped
neutrons can escape a trapping volume before they $\beta-$decay.
Failure to account for this loss mechanism leads to
systematic measurement error and associated 
systematic uncertainties in neutron lifetime 
measurements [4-9].
In this work, we focus on
wall losses of marginally trapped
Ultracold Neutrons (UCNs)
in a
magnetic trapping experiment at NIST.
Since the probability of losing a neutron
when it interacts with materials at a wall
depends, in part, on neutron energy, the survival
probability 
of a marginally trapped UCN
is non-exponential.
We estimate
the systematic error in the neutron lifetime estimate
due to marginally trapped UCNs
for a particular subset of the NIST data where the
trapping potential is static.
We stress that
systematic error (and associated systematic
uncertainty) associated
with marginally trapped UNCs can be reduced
by field ramping strategies to very low levels with
the cost of increasing random uncertainty.
Hence, the systematic error we report for
the static 
potential  case
does not apply to cases
where marginally trapped UCNs are purged by
field ramping methods.

In the NIST experiment [5,6]
an UCN is
produced during a trap-filling stage when a
12 K neutron is scattered to near rest
in liquid helium by single phonon emission [10].
After filling the trap and blocking the cold
beam that produces UCNs, 
light produced by
neutron decay and background events is detected
by a pair of phototubes.
A marginally trapped (or above threshold) neutron is
an
UCN with sufficient energy
to
escape the trap
by interacting with
materials at the  boundary of the trap [11,12,13].
A UCN
with insufficient energy to escape the trap by the wall
loss mechanism is a below threshold UCN.
Since a marginally trapped UCN can be lost to the walls
before it $\beta-$decays,
the decay rate of marginally trapped UCNs is not
exponential.
Hence,
when wall losses are non-negligible,
on average,
the observed $\beta-$decay
rate of neutrons is also non-exponential.
Thus,
if one were to fit an
exponential model to
such decay event data, 
one would
expect
the associated estimate of the neutron lifetime
to be biased.

Here, we develop
a competing risks survival analysis model
[14,15,16,17]
that accounts for
loss of UCNs due to
$\beta-$decay,
wall losses, 
upscattering
and other absorption mechanisms.
Determination of the survival probability function
associated with the wall loss mechanism is the
key problem addressed by this work.
We determine the survival probability function
associated with wall losses
by a Monte Carlo 
method as described in [18].

Based on the conditional survival
probability function of an UCN
that survives the filling stage,
we construct a prediction model for observed
data.
If $\beta-$decay and wall losses
were the only loss mechanisms, and we had perfect
knowledge of the survival probability
of marginally trapped UCNs 
as well as
possible backgrounds,
a fit of this model to data would yield a nearly unbiased estimate of the
true neutron lifetime given a sufficient 
amount of data.
In practice, there are other loss mechanisms including
upscattering and neutron absorption by ${}^{3}$He that affect
the mean lifetime of neutrons in the trap. Still,
given perfect knowledge
of the survival probability functions associated with these
loss mechanisms,
one could
correct the measured
neutron lifetime of trapped UCNs
and recover the true neutron lifetime in an ideal 
noise-free
experiment.
We analyze a subset of
the NIST experimental data
where neutrons are confined in a static potential
to quantify various sources of systematic
measurement error and associated
uncertainty.
For the data analyzed,
the measured lifetime of trapped UCNs is
approximately 700 s.
We attribute the discrepancy between our estimate and the
current best estimate of 880.1 s to
an
additional absorption loss mechanism.
Currently, experiments
are underway to determine if ${}^3$He 
impurities are sufficiently high to explain
the discrepancy.

In some runs of the NIST experiment,
we
vary the trapping potential
to purge marginally trapped
UCNs.
In a simulation experiment,
we demonstrate that field ramping
can in principle reduce systematic measurement error
and associated uncertainty
due to
marginally trapped UCNs to very low levels at the expense
of purging below threshold UCNs.
For a particular example,
we demonstrate that the above
ramping strategy is more efficient than one
where neutrons are trapped when the field is fully ramped
and then increased to its maximum value.
By more efficient, we mean that the expected number of
below threshold UCNs after ramping is higher
for our ramping strategy as compared to the alternative
strategy given that both methods purge marginally trapped
UCNs with nearly the same efficiency.

In this paper, we first
present our physical
and computational methods 
for
simulating UCN trajectories
and loss probabilities of
UCNs due to wall interactions.
Next, we develop our competing risks
survival analysis model
and
apply it to
experimental data acquired at NIST.
We also quantify uncertainties
associated with Monte Carlo sampling
variability and imperfect knowledge of
physical models for neutron interactions with
materials at the walls of trap and beam divergence
effects. 
Finally,
in a Monte Carlo simulation experiment,
we demonstrate
that field ramping
can reduce systematic error due to wall losses of UCN to a very low level
at the expense of reducing the expected number of
trapped UCNs after ramping.


\section{Physical model}
\subsection{Trapping potential model}
In the NIST experiment,
an
UCN
with sufficiently low energy
is
trapped in the potential field
produced 
by
the interaction of the magnetic moment of
a neutron and
a spatially varying magnetic field (Figure 1) 
and gravity. 
For tracking UCNs in the trapping volume,
assuming adiabatic spin transport,
the potential
is
\begin{equation}
V({\bf{x}}) = {\bf {\mu} }  |\B({\bf{x}})|  + m_n g y,
\end{equation}
where $\B$ is the magnetic field, 
$m_n$ and $\mu$ are the mass and magnetic moment
of the neutron respectively, ${\bf{x}}=(x,y,z)$ is spatial location 
of the neutron in the
Cartesian coordinates shown in Figure 1,
and $g$ is the acceleration due to gravity.

We define a nominal trapping volume 
corresponding to $ -z_o \le z \le z_o $ where $z$ is the axial coordinate
and $z_o = $ 37.5 cm (see Figure 1).
An UCN with total energy (kinetic plus potential) greater than the minimum of the potential $V_{\mathrm{min}}$
on the boundary of the nominal trapping volume
is defined to be
a marginally trapped (or above threshold) UCN.
An UCN with energy less than or equal to $V_{\mathrm{min}}$ is
a below threshold
UCN.
We stress that there is not a physical boundary at $z=$ -37.5 cm.
However,
as shown in Figure 1,
as $z$ is reduced from $z \approx$ -37.5 cm to -60 cm,
the trapping potential
decreases dramatically.
For modeling purposes and to speed up Monte Carlo simulations, it is important 
to define a nominal trapping volume so
that all UCN with initial  energy less than
the minimum value trapping potential on the boundary of the nominal trapping
volume
never interact with materials at the ZZ at $z= $ -60 cm and
$z = $ 37.5 cm, nor the cylindrical boundary $r = $ 5.6 cm.
If we choose the lower axial boundary to be much less than
-37.5 cm, this condition is violated.
Different choices of the minimum axial location of the nominal
trapping volume affect results. In Section 4.3, we quantify
this effect.
For a typical static trap configuration of the NIST experiment,
$V_{\mathrm{min}} ~= $ 139 neV.
We assume that 
probability distribution function of the
initial 
speed $| v| $ of an UCN produced in the trap has a quadratic form [18,19]
$f( |v| ) \propto |v|^2 $
for low velocities of interest.

We determine a neutron trajectory based on its 
initial position and momentum,
by solving
the classical equations of motion 
\begin{equation}
\dot{\p} =  F(\bf{x})  = - \nabla V 
(\bf{x}),
\end{equation}
and
\begin{equation}
\dot{\bf{x}} =  \frac{\p}{m_n},
\end{equation}
with an optimal 
fourth order symplectic integration scheme [18,19,20,21].
We predict $|\B|$
at arbitrary points in the trapping volume 
with a three-dimensional tensor-product
spline interpolant [22] where
the order of the spline is four in each direction.
We determine the tensor-product B-spline coefficients 
from 
values of $|\B|$ computed on a grid
by a
code that solves the Biot-Savart law numerically
based on the
geometry of the solenoid and current bars that
produce the magnetic field.
With this tensor-product method,
we evaluate the
potential and its gradient at arbitrary locations
within the trap.

In some experiments, we ramp the quadrupole field
$\B_{q}$
while keeping the solenoid field $\B_{s}$ constant.
Given that the filling stage ends at time $t_L$,
the ramping factor for the quadrupole field is $R(t - t_L)$,
the magnitude of the $\B$-field 
varies as
\begin{eqnarray}
|\B(t-t_L)| = |\B_{s} + R(t-t_L) \B_{q,max}|,
\end{eqnarray}
where
$\B_{q,max}$ is the maximum value of the quadrupole field.
For ramping cases, we develop a tensor-spline method to
model the gradient of $|\B(t-t_L)|$
based on tensor-spline models for each component of
$\B_{s}$ and $\B_{q,max}$.

\subsection{Wall loss model}
When an
above threshold UCN collides with
the cylindrical wall or endcap boundaries,
we assign it a loss probability $p_{{\mathrm{loss}}}$ according to a
model based on assumed material properties [11,12,13].
After $k$ collisions, the empirical survival probability
of the UCN is
\begin{equation}
p_{\mathrm{surv}}(k) = \prod_{i=1}^{k} [ 1 - p_{\mathrm{loss}}(i) ].
\end{equation}
We track each above threshold UCN until
its empirical survival probability drops below $10^{-9}$.
Due to chaotic scattering effects,
the symplectic integration  prediction for a trajectory of an UCN 
does not converge in general as the time step parameter
in the integration code is reduced [19]. Here, we assume
that the mean survival probability
at any time $t$
for an ensemble of
UCNs does not depend on the time step parameter even though
the predicted survival probability of a particular UCN may depend
on the time step parameter.
Although we are unaware of any proof that this
assumption is true,
it seems reasonable. For more discussion of this point,
see [19].

We model the interaction of the neutron
with the materials
on the trap boundaries
with a one-dimensional optical model based
on Schrodinger's equation.
In this approach, we assume that 
each neutron 
energy is sufficiently low 
so that 
its wavelength is very large compared
to the spacing between nuclei
in the wall materials.
Hence,
coherent effects are significant and
interactions are well predicted by an optical model
where the neutron potential is $V - iW$.
For the cylindrical walls,
we model the
the neutron potential
as due to
layers of different homogenous materials
following [11].
The materials at the endcaps at $z=-60$ cm and 
$z= 37.5$ cm are Teflon FEP
\footnote{
Certain materials are identified in this paper to foster understanding. Such identification does not imply recommendation or endorsement by the National Institute of Standards and Technology, nor does it imply that the materials identified are necessarily the best available for the purpose.}
and acrylic respectively.
Given that
a UCN with velocity $\vv$
crosses the trap boundary at location $\x$ and
that
the surface normal for the trap boundary at $\x$ 
is $\nhat$, we define
$E_{\bot} = \frac{1}{2} m_n | \vv \cdot \nhat |^2$.
For the endcaps,
we model the probability of diffuse reflection off the wall based on
Eq. 2.71 of [11] as
\begin{eqnarray}
p_{scat} = \frac
{E_{\bot} - \sqrt{ E_{\bot}( 2 \alpha - 2 ( V_* - E_{\bot} )) } + \alpha}
{E_{\bot} + \sqrt{ E_{\bot} (2 \alpha - 2 ( V_* - E_{\bot} )) } + \alpha},
\end{eqnarray}
where
\begin{eqnarray*}
\alpha = \sqrt{ (V_* - E_{\bot} ) ^2 + W^2 },
\end{eqnarray*}
and
$ V_* = V - V_{He}$
where $V_{He} = $ 15.98 neV.
For the Teflon material, $V=$ 27.8 neV
and
$W=$
1.39e-04 neV.
For the acrylic material, $V=$ 121.04 neV and $W=$
4.74e-05 neV.

The material that coats the cylindrical walls of the trap
is
modeled as multilayer
of 
tetraphenyl butadiene (TPB)
($\mbox{C}_{28}\mbox{H}_{22}$),
Gore-Tex
($\mbox{C}_2\mbox{F}_4$),
graphite ($\mbox{C}$), and
boron nitride ($\mbox{BN}$).
The TPB used in the experiment
is not deuterated and therefore contains a substantial amount of
hydrogen, which has a very large incoherent scattering cross section.
Since
the optical model does not account for incoherent scattering,
we add an additional term to account for it.
The real and imaginary components of the augmented potential for each multilayer are
\[ V = \frac{2 \pi \hbar^2}{m_n}\sum_i
N_i a_i, \]\ and \[ W = \frac{\hbar v}{2} \sum_i N_i \left( \sigma^i_{\mathrm{abs}}
+ \sigma^i_{\mathrm{loss}} \right), \] where for the $i$th nuclear
isotope, $N_i$ is nucleus density;
$a_i$ is the coherent scattering length; $\sigma^i_{\mathrm{abs}}$
is the absorption cross section;
and $\sigma^i_{\mathrm{loss}}$
accounts for incoherent scattering losses.
Since
approximately
half of the neutrons that undergo incoherent scattering will
be scattered back into the ${}^{4}$He and the rest will be lost,
$\sigma_{\mathrm{loss}}$
is set to half of the estimated total incoherent cross section.
After this modification, we solve the appropriate differential
equation and determine the loss probability of the neutron
as a function of $E_{\perp}$ (Figure 2).
Our model does not include the possibility of surface
contamination [23] as we do not currently have a way of characterizing
surface contamination quantitatively.  However, for the most
part, such contamination would lead to an additional marginally-trapped
neutron loss mechanism. We expect that
such an additional loss mechanism
would reduce the systematic effect associated with
marginally trapped neutrons.

After each wall collision, we scatter the neutron back
into the trapping volume.
Since the surface of the walls that scatter neutrons is
rough,
we model the 
reflection
of the neutron 
with 
a Lambertian model [24]
rather than a specular model
that is appropriate for a perfectly smooth surface.
In optical applications of
the Lambertian model,
the intensity of 
a reflected
signal is proportional to 
$\cos( \theta)$
where
$\theta$ is
the inclination angle between the surface normal of the emitter and the direction of the reflected radiation.
Hence, in our simulation
studies,
it follows that
the cumulative distribution function (CDF) for
the inclination angle between the surface normal of the wall and the direction of reflected neutron  $\theta$  is 
\begin{eqnarray*}
F_{L}(\theta) =   \frac{ 1 - \cos( 2 \theta) } { 2 },
\end{eqnarray*}
where $0 \le \theta \le \frac{\pi}{2}$.
The azimuthal angle $\phi$ is a uniformly
distributed random variable between 0 and 2$\pi$.
In contrast, for a diffuse reflection model based on
a uniform intensity model,
the CDF of the direction cosine  of the reflected neutron 
is
\begin{eqnarray*}
F_{U}(\cos(\theta)) =  \cos(\theta),
\end{eqnarray*}
where
$0 \le \cos(\theta) \le 1$.
Later, in Section 4.3, we quantify
the variation of the estimated lifetime of trapped
neutrons for different neutron reflectivity
models.

\section{Survival Analysis Model}
\subsection{Mathematical preliminaries}
In survival analysis,
for any loss mechanism, the loss time $T$ of an object (in our case
a neutron) created at time $t=0$ is
a random variable
with
survival probability $S(t)$ where
\begin{eqnarray*}
S(t) = \Pr ( T > t).
\end{eqnarray*}
For a continuous $S(t)$,
one
can
define
the
hazard function $\lambda(t)$
which
is the instantaneous loss rate 
at time $t$ given that the object of interest survives until time $t$
as follows.
\begin{eqnarray*}
\lambda(t) =
\lim_{ \Delta t \to 0 }
\frac{ 
\Pr( t \le T \le t + \Delta t | T \ge t )
}
{ \Delta t },
\end{eqnarray*}
where
$\Pr( t \le T \le t + \Delta t | T \ge t )$
is the conditional probability that the
loss time $T$ falls in the interval $[t,t+\Delta t]$
given that $T$ is no less than $t$.
Based on the well known conditional probability
equality $\Pr(A|B) = \Pr( A  \cap B ) / \Pr(B)$,
one gets 
the following well known expression for 
the hazard function
\begin{eqnarray*}
\lambda(t) = \frac{1} { S(t) } 
\lim_{ \Delta t \to 0 }
\frac{ 
S(t) - S(t + \Delta t)
}
{ \Delta t }
= - \frac{ \mbox{d} \log{ S(t) } } 
{ \mbox{d} t }.
\end{eqnarray*}

For the neutron $\beta-$decay loss mechanism,
the associated hazard function is $\lambda_{\beta} = \frac{1}{\tau_n}$ where
$\tau_n$
is the neutron lifetime.
For the mechanism associated with
wall losses of marginally
trapped UCNs, we expect $\lambda(t)$
to vary with time since high energy UCNs should, on average,
be lost at a higher rate compared to low energy UCNs.
In general, for any loss mechanism,
\begin{eqnarray*}
S(t) = \exp( -\Lambda (t) ) ,
\end{eqnarray*}
where the cumulative hazard function $\Lambda(t)$ is
\begin{eqnarray*}
\Lambda(t) = 
\int_{t=0}^{t} \lambda(t) \mbox{d}t.
\end{eqnarray*}
Thus, the
survival probability function associated with
$\beta-$decay is $S_{\beta}(t) =
\exp( - \lambda_{\beta}\tau_n )=
\exp( - \frac{t}{\tau_n} ) $.

In a competing risks model, 
the multivariate survival probability function
is $ S(t_1,t_2, \cdots, t_K) $
where
\begin{eqnarray*}
S(t_1,t_2, \cdots, t_K) = \Pr( T_1 \ge t_1, T_2 \ge t_2, \cdots T_K \ge t_K)
.
\end{eqnarray*}
The term $T_i$ is a random variable representing the
loss time associated with the $i$th loss mechanism.
In general,
for $i \ne j$,
the random variables $T_i$ and $T_j$
need not be independent.
The actual loss time of the object of interest
is
$T = \mbox{min} (T_1,T_2, \cdots T_K)$
where $K$ is the number of loss mechanisms.
One can recover the
survival probability
function 
for the $j$th loss mechanism
at time $t$
by evaluating the multivariate one
at $t_j = t$ 
and $ t_i = 0 $ for all other $i \ne j$.
The cause-specific hazard function for
the $j$th loss mechanism is $\lambda_j$
where
\begin{eqnarray}
\lambda_j(t) = -\frac{ \partial \log{S} } { \partial t_j} 
\bigg|_{t_1 = t_2 =\cdots, t_K =  t}.
\end{eqnarray}

For the NIST experiment, we assume that all neutron loss mechanisms
are independent. Given this independence assumption, we can write
$ S(t_1,t_2, \cdots, t_K) = \prod_{j=1}^{K} S_j(t_j)$
where $S_j(t)$ is the survival probability function associated
with the $j$th loss 
mechanism.
Further, for each independent loss mechanism,
the cause-specific loss mechanism for the $j$th loss mechanism is
\begin{eqnarray}
\lambda_j(t) = -\frac{ \mbox{d} \log{S_j}(t) } { \mbox{d} t}.
\end{eqnarray}
For our problem,
the joint hazard 
and survival probability functions of an UCN 
due to all loss mechanisms at
time $t$ are $\sum_{j=1}^{K}\lambda_j(t)$
and
$\exp( 
-\sum_{j=1}^{K} \Lambda_j(t) ).
$ respectively.

\subsection{Conditional survival probablity}
If an UCN is created at time  $t=0$, the conditional
survival probability associated with the first loss mechanism
given the the UCN survives all loss mechanisms  until at least time $t_L$ is
\begin{eqnarray*}
S(t,t_L,t_L,\cdots t_L)  ~/~
S(t_L,t_L,t_L,\cdots t_L),
\end{eqnarray*}
for $t \ge t_L$.
In the NIST experiment,
we model
the creation time of any
UCN during the filling stage as a uniform random variable
between $t=0$ and $t=t_L$ where
$t_L$ is the time spent filling (loading) the trap.
For the first loss mechanism, the conditional survival probability
for an UCN that survives the  filling stage is
\begin{eqnarray*}
S_1(t | T \ge t_L ) = [  \int_{s=0}^{s=t_L} S(t-s,t_L-s,\cdots, t_L-s) ds ] ~/~
\end{eqnarray*}
\begin{eqnarray}
[ \int_{s=0}^{s=t_L} S(t_L-s,t_L-s,\cdots, t_L-s) ds ] 
\end{eqnarray}
A similar statement applies to the other loss mechanisms.
Aside from $\beta-$decay and wall losses,
neutrons can be lost by absorption processes
associated with impurities (primarily ${}^3$He capture )
and by upscattering.
We expect the hazard function associated with upscattering
to vary with temperature.
However, for the ideal case where temperature is 
constant, the hazard function for upscattering is
a constant $\lambda_u = \frac{1}{\tau_u}$.
From first principle arguments,
the hazard function associated with an
energy-independent absorption process is
a constant
$\lambda_a = \frac{1}{\tau_a}$.
We define $\lambda_{*}$ to be sum of the three
hazard functions associated with
$\beta-$decay, upscattering and
an additional
energy-independent
absorption process.
Each of these three hazard functions is a constant in our model.
Thus,
\begin{eqnarray}
\lambda_{*} = \frac{1}{\tau_{*}} =   \frac{1}{\tau_n} + \frac{1}{\tau_u} + \frac{1}{\tau_a}.
\end{eqnarray}
We stress that $\lambda_{*}$ does not account for
the hazard function associated with the wall loss mechanism.
Next, we develop a model that enables us to directly
estimate $\lambda_{*}$ from experimental data given that
we have an estimate for the survival probability function
$S_M(t)$
associated with the wall loss mechanism.

Following arguments in [18],
given the production rate of below and above threshold UCNs
during the filling stage
are $r_{-}$ and $r_{+}$, we
predict the expected number of below threshold UCNs  at the end
of filling stage ($t_L$) as
\begin{eqnarray}
\left< N_{-} (t_L)\right> = r_{-} \int_{s=0}^{t_L} \exp( \lambda_{*} (s-t_L) ) ds.
\end{eqnarray}
The predicted number of above threshold UCNs at the end of the filling stage is

\begin{eqnarray}
\left< N_{+} (t_L)\right> = r_{+} \int_{s=0}^{t_L} S_{M}(t_L-s) \exp( \lambda_{*} (s-t_L) ) ds,
\end{eqnarray}
where
$S_M(t)$ is
survival probability function
associated with wall losses of
above threshold UCNs.
As stated earlier, $S_M(t)$ has a non-exponential
form in general.

Given that an above threshold UCN survives the filling stage,
the conditional survival probability
associated with wall losses is
\begin{eqnarray*}
S^{+}_M(t) = S_M(t | T \ge t_L ) = [  \int_{s=0}^{t_L} S_M(t-s) \exp( \lambda_{*}(s-t_L)) ds ] ~/~
\end{eqnarray*}
\begin{eqnarray}
[  \int_{s=0}^{t_L} S_M(t_L-s) \exp( \lambda_{*}(s-t_L)) ds ] 
.
\end{eqnarray}
For loss mechanisms with exponential survival probability functions,
$S(t | T > t_L) = S(t-t_L)$.
Hence, for times $t > t_L$,
\begin{eqnarray}
\left< N(t) \right> = 
\left< N_{-}(t_L) \right> 
[ ~  1 + \frac{ \left< N_{+}(t_L) \right> }
{ \left< N_{-}(t_L) \right> } 
S^{+}_{M}(t) ~  ]
\exp( -\lambda_{*} (t-t_L) ) 
.
\end{eqnarray}
We can rewrite the above as
\begin{eqnarray}
\left< N(t) \right> = 
\left< N_{-}(t_L) \right> 
[ ~  1 + \Delta(t) ] 
\exp( -\lambda_{*} (t-t_L) ),
\end{eqnarray}
where the time-dependent ``distortion" term $\Delta$ is
\begin{eqnarray}
\Delta(t) = 
\frac{ \left< N_{+}(t_L) \right> }
{ \left< N_{-}(t_L) \right> } 
S^{+}_{M}(t) .
\end{eqnarray}

Based on the above,
the expected loss rate of neutrons lost due to
$\beta-$decay,  absorption, and upscattering are
$r_{\beta}(t)  = \frac{ \left< N(t) \right> } { \tau_n},
r_{a}(t)  = \frac{ \left< N(t) \right> } { \tau_a},$
and
$r_{u}(t)  = \frac{ \left< N(t) \right> } { \tau_u}$ respectively.
Given that upscattering losses are unobservable
and  absorption events yield events that are
indistinguishable from $\beta-$decay events, the overall predicted detection rate
is
\begin{eqnarray}
r_{det}(t)  = \left< N(t) \right>
[  \frac{ p_{\beta}} { \tau_n} +  
\frac{ p_a} { \tau_a}   ] 
\end{eqnarray}
where $p_{\beta}$ and 
$p_{a}$ are detection probabilities for two loss mechanisms.

\subsection{Contamination ratio}

Following [18], we can decompose
$\left< N (t) \right>$
into exponential and non-exponential 
components as follows.
\begin{equation}
\left< N(t) \right>  = 
 f_{exp}(t) + f_c(t),
\end{equation}
where
\begin{equation}
 f_{exp}(t)
=
\left< N_-(t_L) \right> [ 1 + \Delta(t_{end}) ]
\exp( - \lambda_{*} (t-t_L)),
\end{equation}
and
\begin{equation}
f_c(t)
=
\left< N_-(t_L) \right> [ \Delta(t) - \Delta(t_{end}) ]
\exp( - \lambda_{*} (t-t_L)).
\end{equation}
The ratio of the non-exponential and
exponential terms can be
regarded as a contamination ratio $r_c$
where
\begin{equation}
r_c(t)
=\frac
{ \Delta(t) - \Delta(t_{end}) }
{1  +  \Delta(t_{end}) }
.
\end{equation}

We emphasize that $\Delta(t)$ and $f_c(t)$
are nonlinear functions of 
$\lambda_{*}$.
In Section 4,
we estimate
$\lambda_{*}$ 
directly 
from experimental data
given a Monte Carlo estimate of 
$S_M(t)$ based
on a physical 
model for the wall loss probability 
and how surviving
neutrons are reflected back into the trapping volume.
The uncertainty
of $\lambda_{*}$ depends, in part, 
on imperfect knowledge of: the wall loss probability model;
how surviving neutron reflect off the walls;
the spatially varying
fluence of the thermal beam that produces UCN in the trap;
and the usual counting statistics variability in the observed data.

\section{Experimental Application}

\subsection{Prediction model}
For experimental data corresponding
to a particular subset of runs from the NIST experiment,
we estimate $\lambda_{*}$ by fitting a model to experimental data based on Eqns. 15 and
17.
The primary (background plus neutron events)
data are acquired for a static trapping
potential.
Background measurements are also acquired in non-trapping
runs and subtracted from the primary observations [5].
Our primary goal is to understand systematic measurement
errors and associated uncertainties
associated with marginally trapped neutrons and other
loss mechanisms.
Hence, we analyze data corresponding to an experiment
where we did not ramp the magnetic field in order to
maximize the systematic effects of marginally trapped
neutrons.

More specifically,
for bins that are 1 s wide,
we
predict the number of background-corrected counts in 
the $k$th  bin as
\begin{eqnarray}
\hat{n}_k = 
A \lambda_{*} \delta_t [ ~  1 + \Delta(t_k) ] 
\exp( -\lambda_{*} (t_k-t_L) )
 + r_{bg} \delta_t,
\end{eqnarray}
where $\delta_t = $ 1 s
is the resolution at which we determine
survival probabilities by our Monte Carlo method,
$t_k$ is the midpoint of the
$k$th time bin,
and $A$, $r_{bg}$ and  $\lambda_{*}$ are adjustable model parameters determined by
our modeling fitting procedure.
Since the width of the bins for the observed data is 15 s,
we aggregate predictions at the 1 s
scale to get predictions at the 15 s scale of interest.

Given that we estimate these parameters to be
$\hat{A}$ and $\hat{\lambda}$, we can predict
the expected number of below threshold trapped UCNs at $t=t_L$ as
\begin{eqnarray}
\widehat{ \left< N_{-}(t_L) \right> } =
\frac{ \hat{A} \hat{\lambda}_{tot } }
{
p_{\beta} ( \hat{\lambda}_{tot} - \lambda_a - \lambda_u )
+
p_{a} \lambda_a, 
}
\end{eqnarray}
where the hazard functions $\lambda_a$ and
$\lambda_u$  and detection probabilities
$p_a$ and $p_{\beta}$
are
determined from other experiments and/or theoretical arguments.

\subsection{Estimation details}
In the NIST experiment, the cold neutron beam
that produces the UCNs is collimated.
Based on an uncollimated measured beam profile
and knowledge of the geometry of the collimator,
we estimate 
a spatially varying neutron fluence image at the
entrance to the detector
(Figure 3) and an associated probability density function for the intersection
of any neutron trajectory and the plane at $z = -60$ cm.
Based on this probability density function,
we simulate intersection points with the Von Neuman 
rejection sampling method [25,26].
For each intersection point,
we simulate a neutron velocity direction vector $\hat{v}$
that has length 1.
For the case where there is no beam
divergence, $\hat{v}$ is parallel to the axial direction
of the trap.
For the case of non-zero divergence, we simulate $\hat{v}$
such that its direction cosine with respect to the $z$-direction
is uniformly distributed in the interval $(\cos( \theta_{\mathrm{max}} ), 1)$.
That is, 
simulated realization of $\hat{v}$
fall within a cone.
The location of the UCN produced by a neutron is
$(x_o,y_o,z_o) + L_{sim} \hat{v} $
where
$(x_o,y_o,z_o)$
is the simulated location of the neutron at $z_o=$ -60 cm,
and $L_{sim}$ is the simulated distance traveled by the neutron
before a UCN is produced.
Given the initial location of the UCN, we
simulate its initial velocity 
as described earlier.
Given the initial velocity and position of an UCN, we determine its
trajectory
with the symplectic integration method described in
Section 2.1.

For the case of  no beam divergence,
the  production rate of
above threshold UCNs, $r_+$ (Eq. 12), produced in the trap
at random times in the interval $(0,t_L)$
is  4.2 times larger
than the production rate, $r_-$ (Eq. 11), of below threshold UCNs
for the nominal trapping volume defined by
-37.5 cm $ \le $ 37.5 cm (Figure 4).
The energy range of
simulated above threshold UCNs
is
139 neV to  246 neV.
In general, the UCN wall collision rate
increases with energy (Figure 5).
Based on Monte Carlo estimates of $S_M(t)$
at discrete times
$(0,1,2,\cdots, 5500)$ s,
we determine $\Delta(t)$ (Eq. 16) 
on a discrete time grid 
(Figure 6).
Recall, the theoretical value of $\Delta(t_L)$ equals
$\frac{ \left< N_+(t_L) \right> } {\left< N_-(t_L)\right> }$.
Even though the relative production of above and below threshold UCNs
is 4.2, 
$\frac{ \left< N_{+}(t_L) \right> } { \left<N_{-}(t_L)\right>} $
is approximately 0.41.
That is, a large fraction of above threshold UCN is lost
during the filling stage.
Since
UCNs with energy greater than 246 neV
would be lost to the walls in approximately a few seconds or less,
such high energy UCNs would have a negligible effect 
on $\Delta(t)$ for $t - t_L > $ 10 s.
Thus, extending the maximum energy of simulated above threshold
UCN would have a  negligible
effect on our estimate of $\lambda_{*}$.

We average 40 independent estimates of $S_M(t)$ from
independent Monte Carlo experiments to get an overall
estimate of $S_M(t)$ and $\Delta(t)$.
Given
a wall loss
probability model that accounts for
incoherent scattering, we determine the mean lifetime
of the neutron in the trap to be 700 s 
with a standard uncertainty
of
57 s (Table 1, Figure 7).
In this standard approach,
the model is assumed to be valid and
deviations between
observations and predicted values based on perfect
knowledge of the model parameters
are due to counting statistics.
To quantify the component of uncertainty due to imperfect knowledge
of $S_M(t)$ due to sampling variability, we 
simulated bootstrap [27] replications of our estimate
of $S_M(t)$
by resampling with replacement the 40 independent estimates
of $S_M(t)$ that we averaged in the first study.
For each bootstrap replication of the average $S_M(t)$, we refit
our model to the same observed data.
The bootstrap estimate of uncertainty due to
sampling variability of
$S_M(t)$ is 2.4 s.

\subsection{Systematic effects}
For comparison, when we set $\Delta(t)=0$, i.e., when we neglect
wall losses, we estimate the lifetime to be 655 s
with an estimated uncertainty of
51 s
(Table 1).
When we estimate $S_M(t)$ based on
a wall loss model that neglects incoherent scattering
effect for the cylindrical boundary,
we estimate the neutron lifetime to be 
708 s with an associated 1-sigma uncertainty of
59 s.
Based on this, we estimate the component of uncertainty
contributed by
imperfect
knowledge of the wall loss probability model
to be 8 s.

For the primary wall loss model that accounts for
incoherent scattering,
we estimate $\lambda_{*}$ for
different definitions of the
trapping volume. In particular,
we vary the lower axial boundary of the trapping volume,
$z_{min}$,
from -37.5 cm to -35 cm 
and -40 cm
(Table 2).
We fit a linear model to predict the expected value of  $\hat{\lambda}$
as a function of $z_{min}$. The estimated slope of this linear model
and its associated uncertainty are -0.55 s/cm
and 0.40 s/cm.

In a similar study,
we vary the time-step parameter in the symplectic
integration algorithm
(Table 2).
The estimated slope of a linear model to predict 
the lifetime as a function of the time step is
2.9 $\times 10^{-4}$ where the uncertainty is
3.79 $ \times 10^{-4}$.
Based on this result, the expected difference in
a lifetime estimate based on a simulation where the time step parameter $\mbox{d}t = $ 
5.0$ \times 10^{-5}$
and where
$\mbox{d}t = \epsilon $ 
where $\epsilon$ is arbitrarily close to 0 s, is
1.4 (1.9) s.

Variation in
the beam divergence parameter $\theta_{max}$
produces
a
statistically significant variation in the lifetime
estimate (Figure 8) because
the distribution of
the initial
potential energy of an UCN created in the trap depends on $\theta_{max}$.
Thus,
even though
the distribution of the initial kinetic energy of an UCN
produced in the trap does not depend on $\theta_{max}$,
the distribution of the initial total energy of the UCN depends on
$\theta_{max}$.
We estimate $\theta_{max}$ to be 3.0 degrees by matching the
expected value of $\theta$ in the simulation to
that predicted
based on analysis of the dispersion of
scattering plane orientations in mosaic
crystals relative to their mean value [28].
In [28], the angle between the random orientation of
a particular scattering plane and the mean orientation is 
a truncated Gaussian.
Since the slope of the fitted line in Figure 8 is
$-$0.97(0.16) s  deg${}^{-1}$, we estimate the systematic error
due to ignoring beam divergence effects to be approximately
2.9 s.
As a caveat,
to estimate an uncertainty,
we assume 
a one-to-one
relationship
between
the 
standard deviation of the random
mosaic crystal orientations
and the standard deviation of 
direction cosines in our simulation study.
Further, in our beam divergence study,
the 
probability distribution function for the direction cosine of
a UCN at $z=$ -60 cm is a uniform distribution.
In contrast, for the mosaic crystal model, the angle
between the mean orientation and any random orientation is
a truncated Gaussian distribution.

Our estimate of $\lambda_{*}$
systematically
depends on the assumed model for how surviving neutrons are reflected
back into the trapping volume (Table 3).
Because of surface roughness effects,
the two diffuse models are much more plausible than
than the specular model. Hence,
the difference between the estimates 
for the diffuse models ( 1.3 s  with an uncertainty
of 1.5 s ) is relevant for estimation of a  systematic uncertainty
due to imperfect knowledge of the model for  neutron reflection.

In all earlier studies in this work, we account for
gravity and spatial variation in the cold beam fluence.
Failure to account for gravity shifts the
estimate 
of $\tau_{*}$ downward by 5.1 s (standard uncertainty is 1.6 s)
(Table 4).
We expect this shift for two reasons. First, gravity
changes the distribution of the initial potential energy of
UCNs and hence the distribution of the initial energy of
UCNs.
Second, for the cases studied here,
gravity reduces the minimum potential energy
on the nominal trapping volume boundary from 139 neV to 
135.5 neV. Recall, this minimum potential energy
is the threshold for defining a marginally trapped UCN.

Failure to account for spatial variation in assumed neutron
beam fluence
shifts the
estimate 
of $\tau_{*}$ upward by 7.5 s (standard uncertainty is 1.6 s)
(Table 4).
As a summary, we list estimates of systematic
uncertainties associated with the effects studied here
in Table 5.

\begin{table}
Table 1.
Model parameter estimates for NIST experimental data
for a static potential.
Survival probability function
associated with wall loss determined for two scenarios;
incoherent scattering 
in neutrons in materials is either
neglected or accounted for.
Beam divergence neglected.
The time step parameter for all cases is ${\mbox{d}t}=$ 5.0e-05 s.
\begin{center}
\begin{tabular}{ccc|ccccc} \hline\hline
\\
case & account for & account for & A &  $\tau_{*}$ (s) &  $ r_{bg}$ (s${}^{-1}$) & $\chi^2/df$ &  p-value
\\
 & incoherent& marginally& &  &  &  & 
\\
 &scattering &trapped neutrons& &  &  &  & 
\\
a & yes & yes &  57943 (3883) & 700 (57)  & 1.8 (1.8)  & 1.080 & 0.2234 
\\
b & yes & no &  76114 (4531) & 655 (51)  & 2.3 (1.7) & 1.085 & 0.2118
\\
c & no & yes &  41558 (2856) & 708 (59)  & 1.8 (1.8)  & 1.082 & 0.2187
\\
\hline
\end{tabular}
\end{center}
\end{table}

\begin{table}
Table 2.
For a static potential experiment,
dependence of estimated $\tau_{*}$
on lower axial boundary of
nominal trapping volume ($z_{min}$) and time-step
$\mbox{d}t$ in symplectic integration algorithm.
We account for wall losses when estimating
$\tau_{*}$.
Beam divergence effects neglected.
Reported uncertainties account for
uncertainty 
of $\Delta(t)$ (Eq. 16) due to Monte Carlo sampling variability.
\begin{center}
\begin{tabular}{ccc} \hline\hline
\\
${\mbox{d}t}$ (s) & $z_{min}$ (cm) &  $ \tau_{*} (s)$
\\
\\
\\
1.0e-04 & -40  & 703.3(1.5)
\\
1.0e-04 &-35 &  700.5(1.5)
\\
1.0e-04 &-37.5 &  701.9(1.1)
\\
5.0e-05 &-37.5 &  699.9(2.4) 
\\
2.5e-05 &-37.5 &  699.9(2.8)
\\
\hline
\end{tabular}
\end{center}
\end{table}

\begin{table}
Table 3.
{For a static potential experiment,}
dependence of estimated $\tau_{*}$
on neutron scattering modeling.
Beam divergence neglected.
Reported uncertainties account for uncertainty
of $\Delta(t)$ (Eq. 16) due to Monte Carlo sampling variability.
\begin{center}
\begin{tabular}{cccc} \hline\hline
\\
\\reflection model& ${\mbox{d}t}$ (s) & $z_{min}$ (cm) &  $ \tau_{*} (s)$
\\
\\
diffuse Lambertian & 1.0e-04 &-37.5 &  701.9(1.1)
\\
diffuse uniform & 1.0e-04 &-37.5 &  703.2(1.0)
\\
specular & 1.0e-04 &-37.5 &  713.5(1.5)
\\
\\
\hline
\end{tabular}
\end{center}
\end{table}

\begin{table}
Table 4.
For a static potential experiment,
dependence of estimated $\tau_{*}$.
on gravity and spatial variation of neutron beam (Figure 2).
For all cases, the wall loss probability model
accounts for incoherent scattering.
We assume a diffuse Lambertial model for neutron reflections.
No beam divergence.
The time step parameter is $\mbox{d}t=$1.0e-04.
The uncertainties account for imprecise
knowledge of the $\Delta(t)$ (Eq. 16)
due to Monte Carlo sampling variability.
\begin{center}
\begin{tabular}{ccc} \hline\hline
\\
gravity accounted for & beam profile accounted for & $ \tau_{*} (s)$
\\
yes &   yes  &  701.9(1.1)
\\
no &   yes  &  694.4(1.2)
\\
yes &   no  &  707.0(1.2)
\\
\\
\hline
\end{tabular}
\end{center}
\end{table}

\begin{table}
Table 5.
For a static potential experiment,
systematic effects and associated uncertainties 
for
$\tau_{*}$ measurement
due to wall losses of marginally trapped
UCNs.
Effects that are not statistically
significant are indicated with asterisks.
\begin{center}
\begin{tabular}{ccc} \hline\hline
\\
effect & correction & uncertainty 
\\
\\
wall loss model& none &  8 s
\\
neutron reflections model& none &  1.3 s
\\
beam divergence& -2.9  & 2.9 s
\\
beam profile model& none & NA
\\
upscattering & none &  $<<$ 1 s
\\
${}^{3}$He absorption & NA & NA
\\
mechanical vibrations & none & NA
\\
time step & none & 1.9 s
${}^{*}$
\\ 
choice of $z_{min}$ & none & 1 s 
${}^{*}$
\\
\hline
Total & -2.9 s &  8.9 s 
\\
\hline
\hline
\end{tabular}
\end{center}
\end{table}

For the NIST experiment,
we expect the walls of the trap
to vibrate and perturb the energy
of
UCNs that survive wall collisions and are reflected back
into the trapping volume.
For more discussion of this effect 
for a simplified 
1-D vibrational model 
for a magneto-gravitational
trap, see Salvat and Walstrom [29].
Since the wall loss probability 
of an UCN depends on its $E_{\perp}$ value,
energy perturbations
would affect the survival probability of above
threshold UCNs for a static trap experiment.
Energy perturbations due to mechanical vibrations
would also affect how well above threshold UCNs
are purged (and how well below threshold
UCNs are retained) when
magnetic fields are ramped.
Development 
of 
a realistic three-dimensional model for the effect
of mechanical vibrations 
for the NIST experiment
is an important and very challenging research topic
beyond the scope of this study.

\section{Ramping Studies}
In magnetic trapping experiments,
one can
purge above threshold UCNs from the trap
by ramping the trapping potential down and then
ramping it back up to its original value [4,30].
In a background-free simulation experiment,
we reduced the quadrupole field 
from its maximum value to an adjustable
fraction of its initial value  (Figure 9). 
We denote this fraction as a ramping fraction.
For any given ramping schedule, we determined
the conditional survival probability
for all UCNs (rather than just above threshold UCNs)
that survive ramping
by a direct simulation method 
for a fixed value of $\tau_{*} = $ 686 s.
Based on this conditional survival probability, we
simulate high-count $\beta-$decay data under the assumption that
the observation rate in narrow 1 s width bins is well approximated as
$\left< N(t) \right> / \tau_n $.
We fit the Eq. 22  prediction model to simulated data with
$\Delta(t) = 0$. That is, we ignore wall losses.
For ramping fractions less than approximately 0.3,
the systematic error of the estimated lifetime is very small
(Figure 10).

The benefits of purging above threshold UCNs by field ramping
come with a cost.
That is, we also purge a non-negligible fraction of
initially below threshold UCNs.
For instance, for the case where the ramping fraction is
0.3, the fraction of surviving above threshold UCNs
immediately after ramping ends is
0.00072. However, the fraction of originally below
threshold UCNs is reduced to 0.30(0.02).

There are other field ramping strategies to purge
above threshold UCNs than the one considered here. For instance, one
could fill the trap with the trapping potential reduced to its
minimal value and then ramp it after the filling stage ends.
In a simulation experiment, we compared this alternative ramping
strategy to ours. We simulated UCNs with the same
initial locations and initial  velocities and tracked them for
both strategies. In order to purge all but 0.0072 of the 
above threshold UCNs, the fraction of
below threshold UCNs retained by the alternative upramping
scheme is 0.06(0.01).
Thus, for this case, the relative number of below threshold UCNs
after ramping is less for the alternative ramping strategy
compared to the ramping strategy implemented in our simulation study.
To help explain the result,
we note that in the alternative ramping schedule,
the quadrupole field is maintained at a much lower initial
value than in the original ramping schedule.
Hence, during the filling stage,
the energy threshold that defines a marginally trapped UCN
is lower for the alternative ramping scheme
compared to the original scheme.

\section{Summary}
We developed a competing risks
survival analysis model to account for
losses due to the joint effect of $\beta-$decay losses,
wall losses of
marginal trapped neutrons, and an unspecified absorption mechanism.
We
determined the
survival probability function associated
with the wall loss mechanism 
by a Monte Carlo method based on physical models
for the loss probability of above threshold UCNs that 
interact with  materials in cylindrical wall and endcaps
of the trap.
In our approach, we track above threshold UCNs in
the trap with a computer intensive symplectic integration method.
For any trapping volume, we objectively define above threshold
UCNs as those with total energy greater than the minimum potential
energy on the trap boundary.
Hence, there is no need to track UCNs below this threshold.

Based on estimated survival probabilities, we constructed a prediction model
for observed data acquired in a magnetic trapping
neutron lifetime experiment at NIST
where the trapping potential is static.
Based on a
fit of this model to a subset of the NIST experimental data,
we determined the mean lifetime of neutrons in the trap
to be
700 s with an uncertainty of approximately 60 s --
considerably less than the current best estimate
of (880.1 $\pm$ 1.1 ) s.
The source of this discrepancy is an ongoing research topic;
experiments are underway to determine if
this discrepancy can be explained by
neutron capture by
${}^3$He impurities in the trapping volume.
In our model,
the largest source of systematic uncertainty
is associated with imperfect
knowledge of the wall loss probability model (8 s) (Table 5).
If we neglect the effect of marginally trapped neutrons,
for the static case considered here, the estimated lifetime
of the neutron is shifted downward by 45 s (Tables 1).

In a Monte Carlo experiment,
we demonstrated that
when the trapping potential is ramped down and
then back again,
systematic error due to wall losses of marginally trapped UCNs
can be suppressed to a
very low level.
For a particular case,  Monte Carlo
simulation studies showed that 
this ramping strategy is more efficient than one
where neutrons are trapped when the field is fully ramped
and then increased to its maximum value.
From a practical perspective, our stochastic model
and associated Monte Carlo
methods should be helpful to guide design of field
ramping strategies to purge above threshold UCNs
in magnetic trapping and related trapping experiments.

\newpage{}
\noindent{}
{\bf Acknowledgements.}
We
thank Grace Yang for useful comments.
We acknowledge the support of the NIST, US Department of Commerce, in
providing support, including the neutron facilities used in this work.
This work is also supported in part by the US National Science
Foundation under Grant No.\ PHY-0855593 and the US Department of
Energy under Grant No.\ DE-FG02-97ER41042.
Contributions by staff of NIST, an agency of the US government,
are not subject to copyright in the US.

\newpage{}


\begin{figure}
\centering
\includegraphics[width=6.0in]{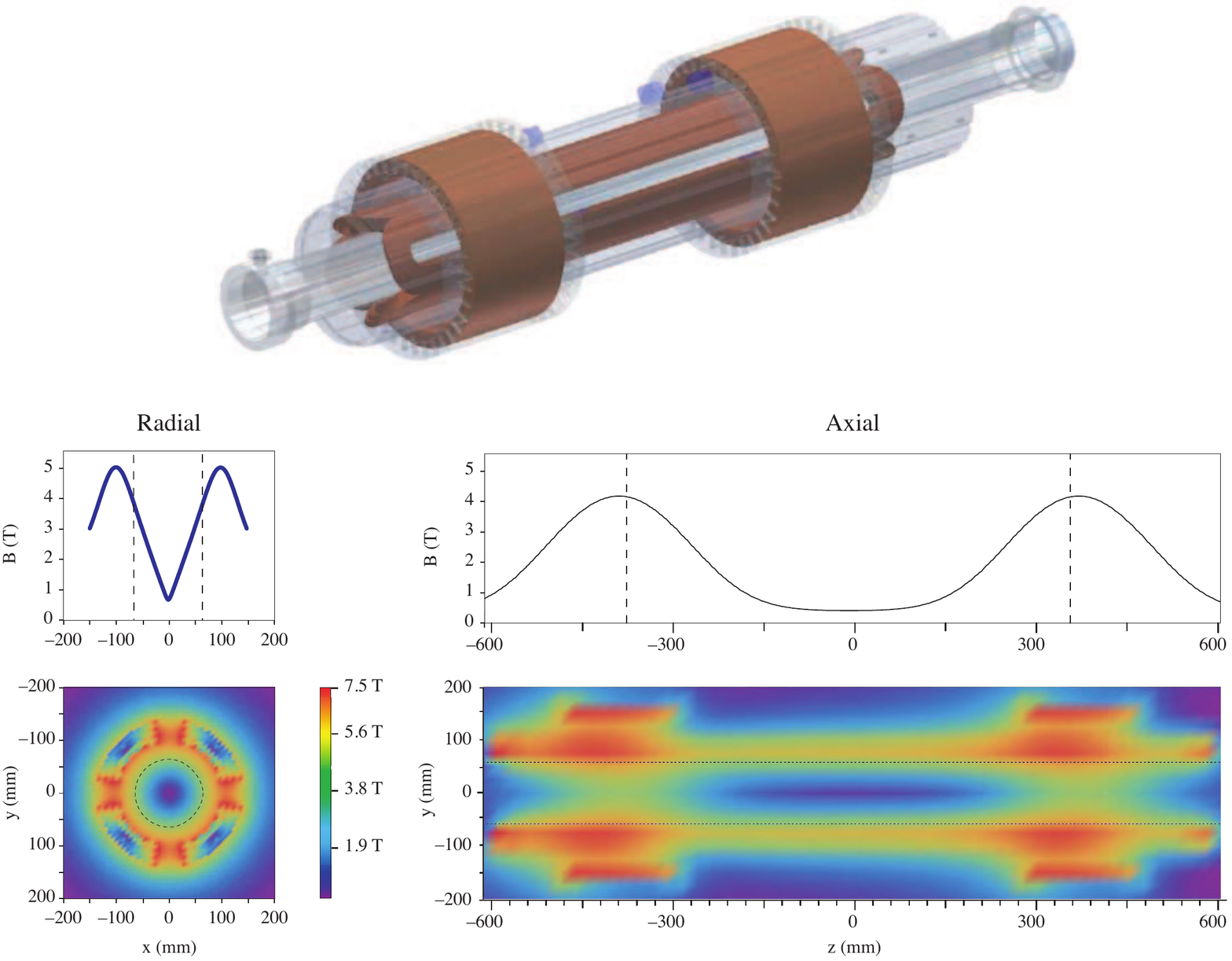}
\caption{
Magnetic field profiles in the
NIST magnetic trapping expeirment.
}
\label{fig_1}
\end{figure}

\begin{figure}
\centering
\includegraphics[width=6.0in,angle=0]{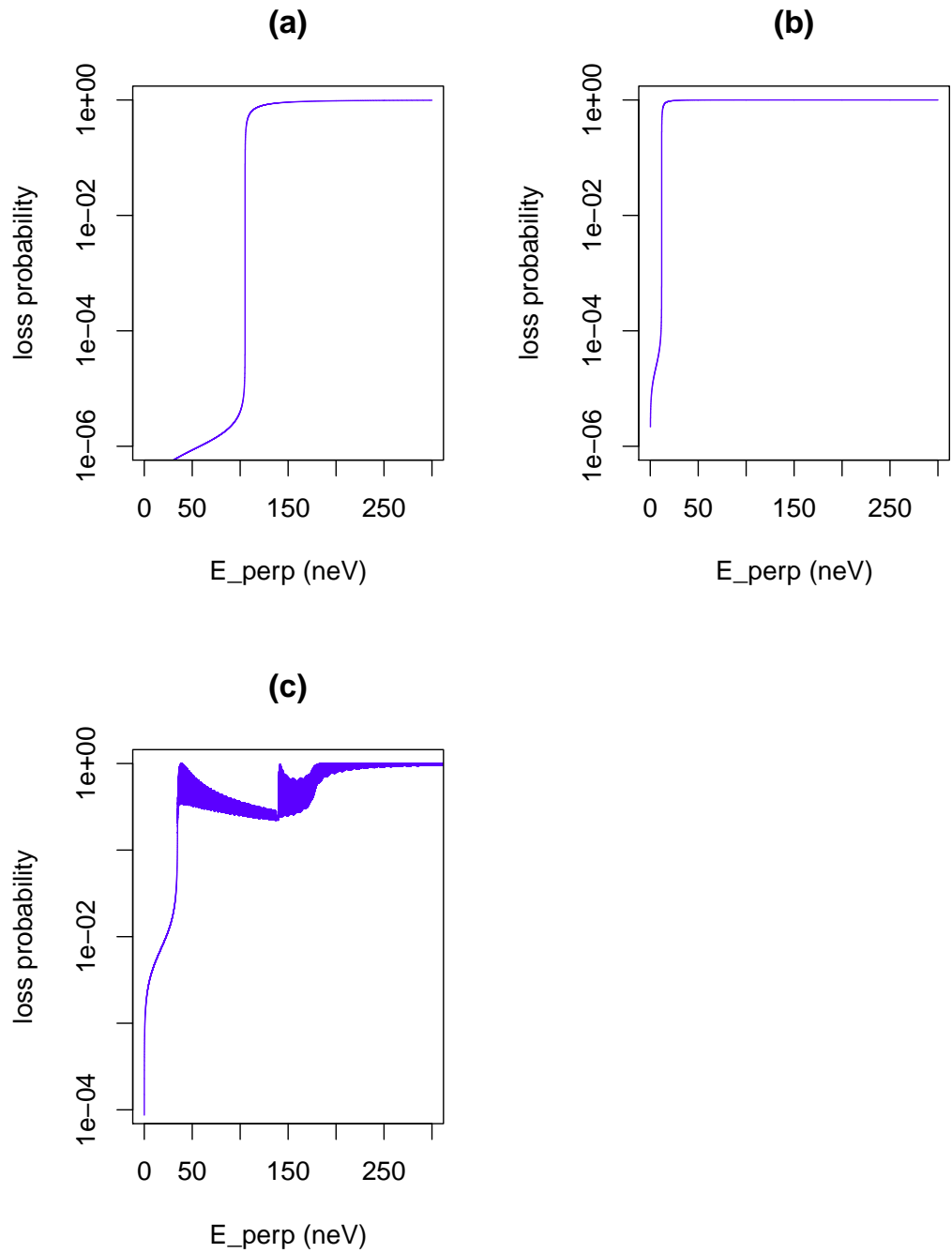}
\caption{
Loss probabilities for (a) the Teflon coated endcap;
(b) the acrylic coated endcap
and
(c) multilayers at cylindrical wall.
}
\label{fig_1}
\end{figure}

\begin{figure}
\centering
\includegraphics[width=4.0in]{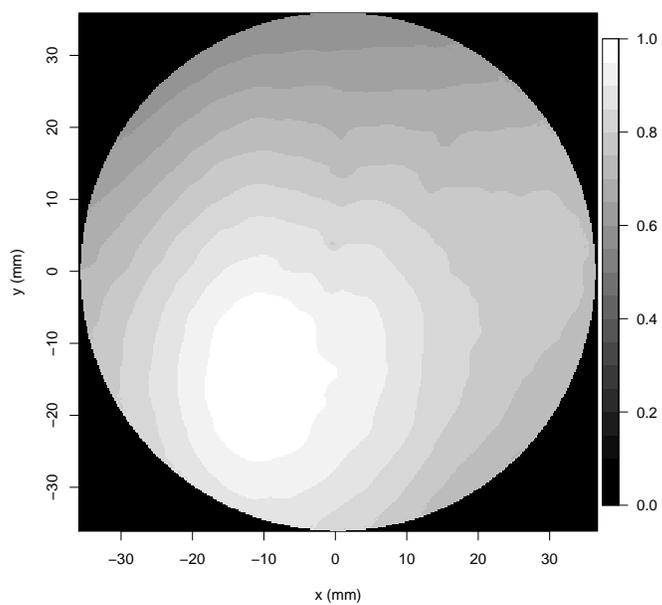}
\caption{Model for
neutron fluence in NIST magnetic trapping experiment
at trap entrance.
}
\label{fig_1}
\end{figure}

\begin{figure}
\centering
\includegraphics[height=5.0in]{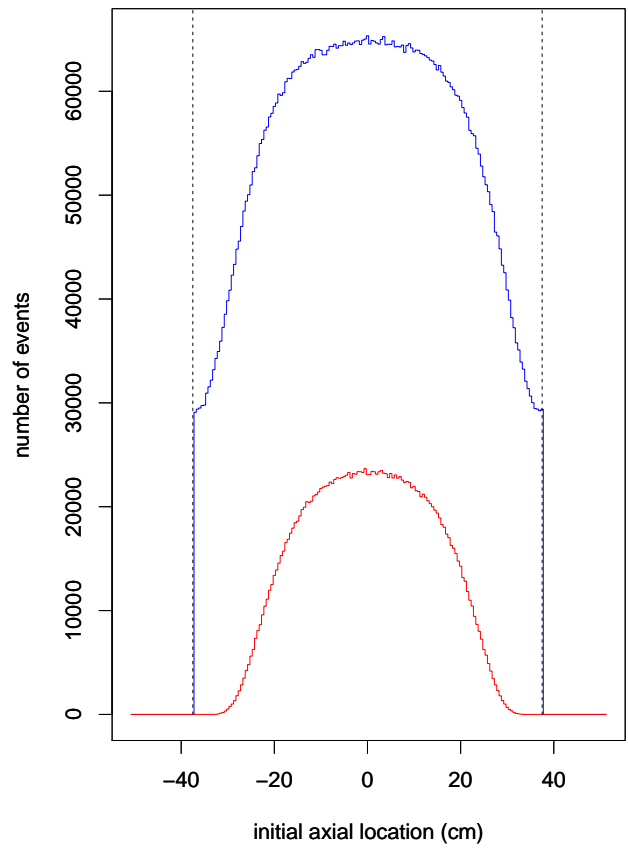}
\caption{
For a static trap where the trapping volume is defined to
be -37.5 cm $ \le z \le$ 37.5 cm, the ratio of above threshold
to below theshold UCNs is  4.22 for energies between
approximately 139 neV to 246 neV.
}
\label{fig_2}
\end{figure}

\begin{figure}
\centering
\includegraphics[width=6.0in]{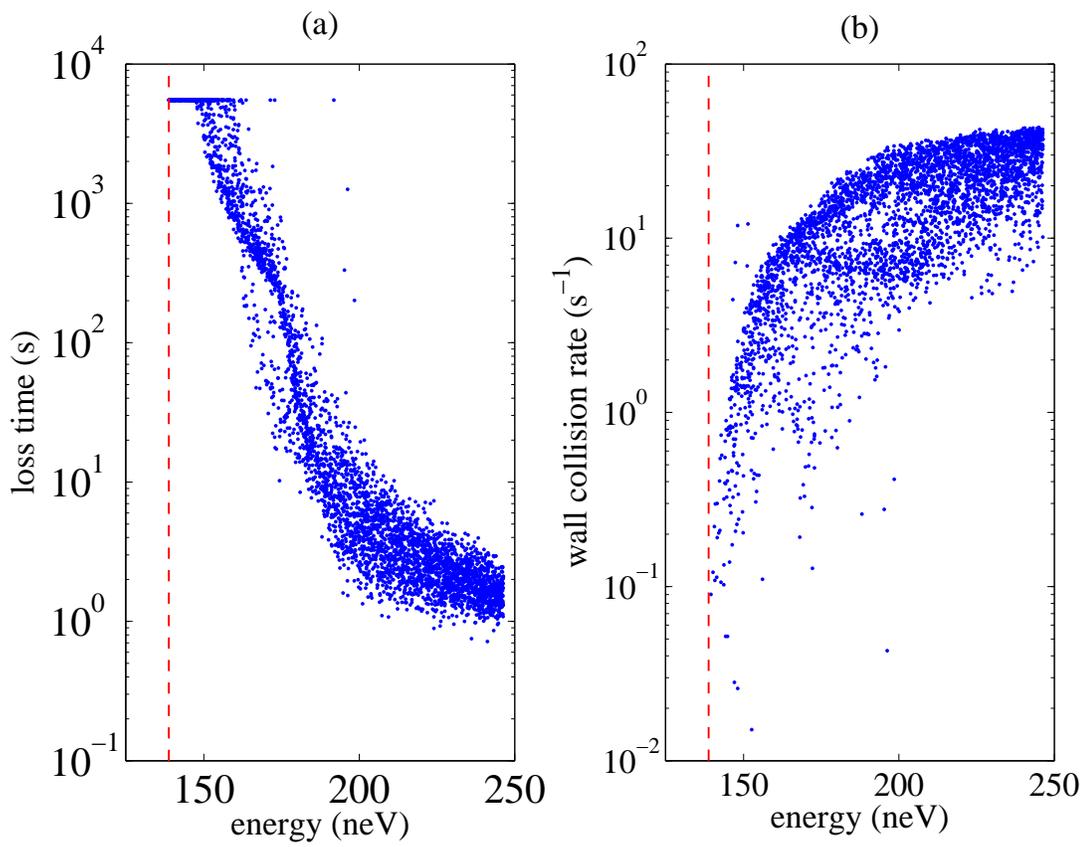}
\caption{
Loss time and wall collision rates for simulated above threshold UCNs.
A UCN is lost when its empirical survival probability falls below
1.0e-09.
Tracking halted 5500 s after creation time 
if a UCN has survival probability above 1.0e-09.
}
\label{fig_2}
\end{figure}

\begin{figure}
\centering
\includegraphics[width=6.0in]{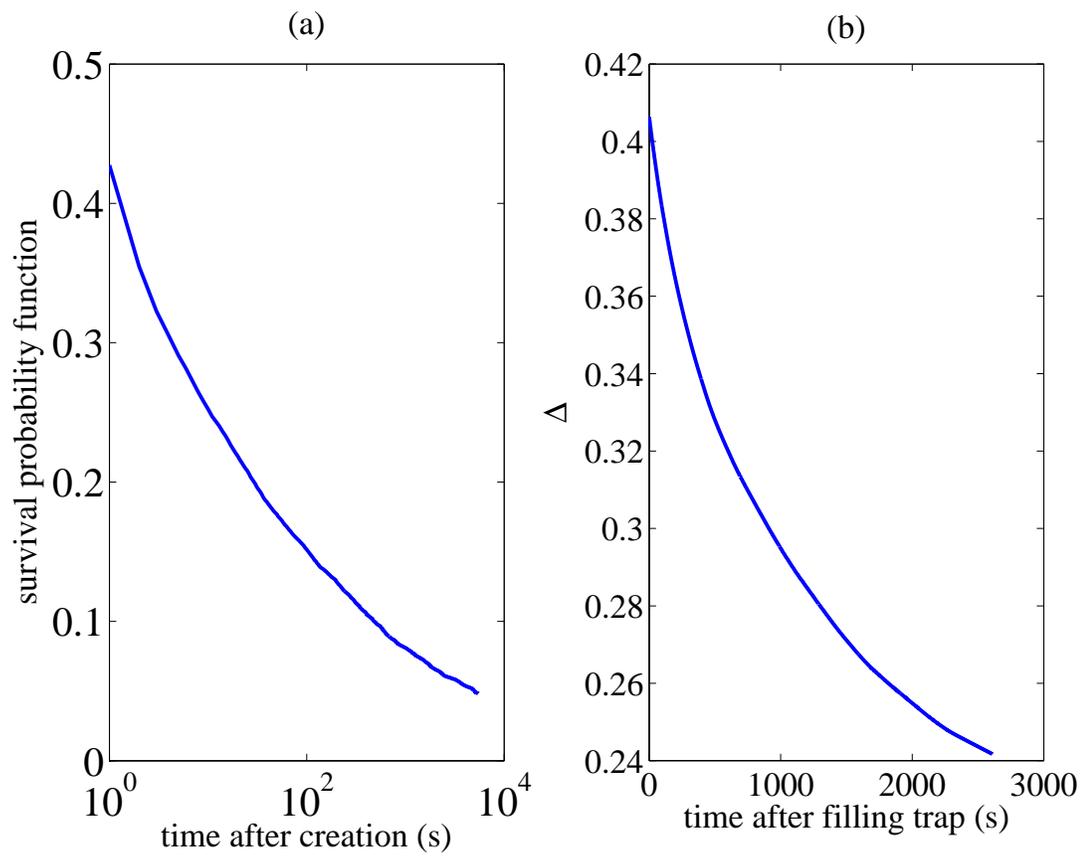}
\caption{
(a). Monte Carlo estimate of survival probability function of
above theshold UCNs.
(b). Monte Carlo estimate of distortion term $\Delta(t)$
(Eq. 16).
}
\label{fig_2}
\end{figure}

\begin{figure}
\centering
\includegraphics[width=4.0in]{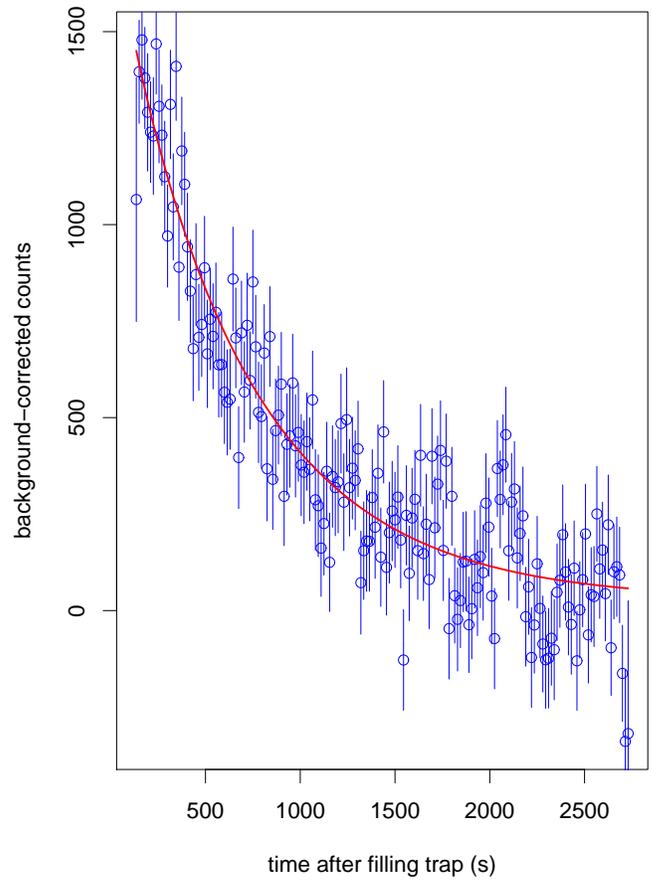}
\caption{
Observed and predicted (line) background-corrected data for
a subset of data from the
NIST magnetic trapping
experiment. 
For this subset, the trapping potential 
is static.
Prediction model accounts for wall losses.
}
\label{fig_3}
\end{figure}

\begin{figure}
\centering
\includegraphics[width=4.0in]{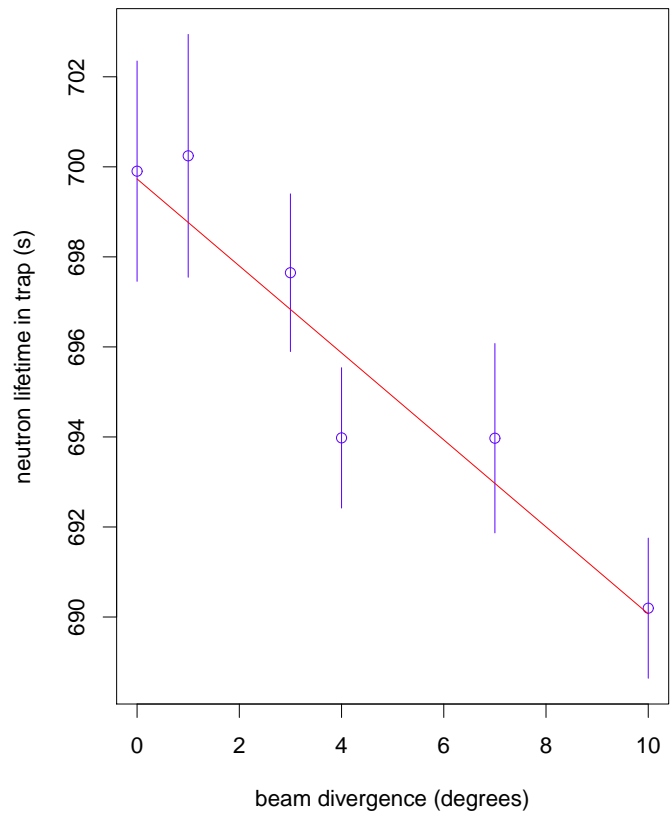}
\caption{Model Carlo estimates of $\Delta(t)$ depend on assumed beam
divergence witin the trap.
Fitted intercept and slope are 699.73(1.03) 
and
$-$0.97(0.17) s / degree.
}
\label{fig_1}
\end{figure}

\begin{figure}
\centering
\includegraphics[width=4.0in,angle=-0]{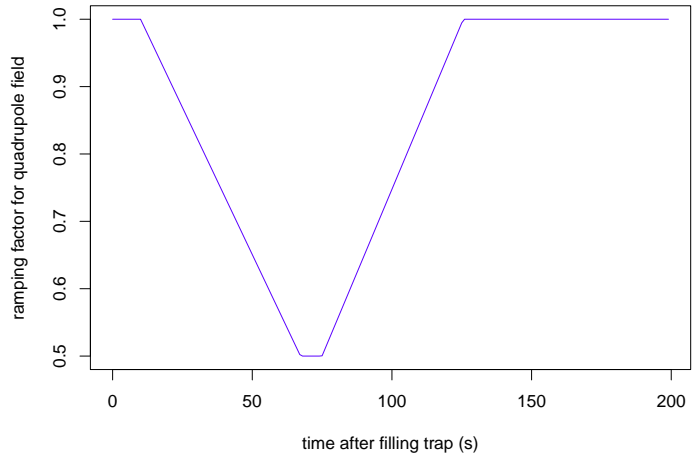}
\caption{In simulation experiment, the quadrupole field
is reduced by a fraction that varies from 1 to an adjustable
minimum.
}
\label{fig_1}
\end{figure}

\begin{figure}
\centering
\includegraphics[width=4.0in,angle=-0]{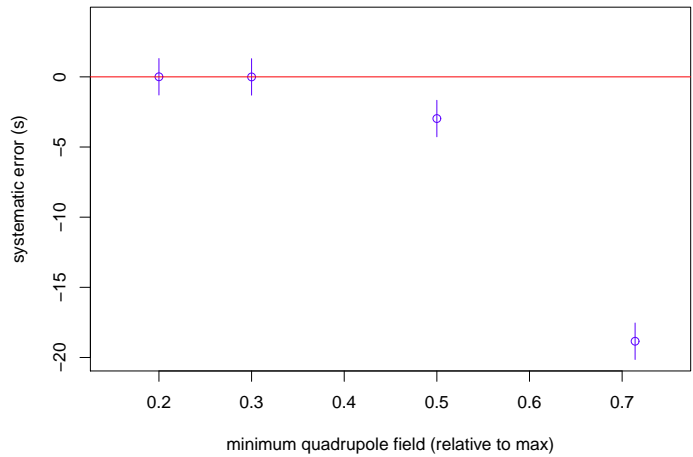}
\caption{Bias of neutron lifetime
when an exponential model is fit to simulated data contaminated by
above threshold UCNs.
For unramped field, bias is $-$31(2) s.
The true value of $\tau_{*}$ is 686 s.
}
\label{fig_1}
\end{figure}

\end{document}